# LitSumm: Large language models for literature summarisation of non-coding RNAs

Andrew Green[1], Carlos Ribas[1], Nancy Ontiveros-Palacios[1], Sam Griffiths-Jones[2], Anton I. Petrov[3], Alex Bateman[1] and Blake Sweeney[1*]

1. European Molecular Biology Laboratory, European Bioinformatics Institute (EMBL-EBI), Wellcome Genome Campus, Hinxton. CB10 1SD. UK
2. Faculty of Biology, Medicine and Health, The University of Manchester, Manchester, M13 9PT, UK
3. Riboscope Ltd, 23 King St, Cambridge. CB1 1AH. UK

*To whom correspondence should be addressed.

## Abstract

Curation of literature in life sciences is a growing challenge. The continued increase in the rate of publication, coupled with the relatively fixed number of curators worldwide presents a major challenge to developers of biomedical knowledgebases. Very few knowledgebases have resources to scale to the whole relevant literature and all have to prioritise their efforts.

In this work, we take a first step to alleviating the lack of curator time in RNA science by generating summaries of literature for non-coding RNAs using large language models (LLMs). We demonstrate that high-quality, factually accurate summaries with accurate references can be automatically generated from the literature using a commercial LLM and a chain of prompts and checks. Manual assessment was carried out for a subset of summaries, with the majority being rated extremely high quality.

We apply our tool to a selection of over 4,600 ncRNAs and make the generated summaries available via the RNAcentral resource. We conclude that automated literature summarization is feasible with the current generation of LLMs, provided careful prompting and automated checking are applied.

**Availability**

Code used to produce these summaries is available here: https://github.com/RNAcentral/litscan-summarization and the dataset of contexts and summaries is available here: https://huggingface.co/datasets/RNAcentral/litsumm-v1.5. Summaries are also available on the RNA report pages in RNAcentral (https://rnacentral.org/)

**Database URL:** https://rnacentral.org/

# Introduction

Curation in life sciences is the process by which facts about a biological entity or process are extracted from the scientific literature, collated and organised into a structured form for storage in a database. This knowledge can then be more easily understood, compared and computed upon. The curation task is a time consuming and often challenging one in which subject matter experts triage literature, select curateable papers and review them for the rich information they provide about a given biological entity (1). Researchers search curated databases (knowledgebases) for information about the entities they are studying and incorporate curated facts into the design of their next study, which may in turn be curated. This virtuous circle is fundamental to the functioning of research in life sciences.

One of the most basic requirements for a researcher is a broad understanding of their molecule of interest. A broad overview is most easily gained from a short summary of the literature. Such summaries are often produced as part of the curation process, for example UniProt (2) gives an overview of a protein's function on its protein entry pages. Similarly, some Model Organism Databases have curator-written descriptions of the genes they contain (e.g. Saccharomyces Genome Database (3) and FlyBase (4)). Summaries are time consuming to produce because there may be a large amount of disparate information to synthesise; because of the difficulty, many databases still do not yet have summaries for all the entities they contain, e.g. the RNAcentral database does not contain summaries for non-coding RNAs (ncRNAs). In addition, human written summaries are prone to become out-dated due to the lack of available curator time.

There are a limited number of curators in the world and the rate of publication and the complexity of the research papers continues to increase. The mismatch between the effort that is required and that which can be applied has led many to use computational techniques at all stages of curation. Natural Language Processing (NLP) has been applied for many years, with cutting edge techniques being used as they become available. However to date these approaches have had limited success. Recently, language models, and in particular Large Language Models (LLMs) have attained sufficient quality to be applicable to curation. Recent efforts have used LLMs to summarise gene sets (5), mine knowledge from synthetic biology literature (6), and other tasks previously done by NLP methods (7). In most cases, LLMs are able to perform remarkably well with little or no fine-tuning training data, opening the potential for their application in resource limited fields.

One field in which the lack of curation effort is particularly acute is ncRNA science. ncRNAs are any RNA transcribed in the cell which does not encode a protein. ncRNAs are critical to the functioning of the cell by forming the core of the ribosome, splicing pre-mRNAs in the spliceosome, and regulating gene expression through microRNAs (miRNAs), long non-coding RNAs (lncRNAs), small nucleolar RNAs (snoRNAs) and many other RNA types. However, as a field, ncRNA has very little curation resource compared to the field of proteins. Rfam (8) and RNAcentral (9) are two of the primary databases in RNA science. Rfam is a database containing over 4,100 RNA families, while RNAcentral is the ncRNA equivalent of UniProt containing over 30 million sequences at the time of writing. Rfam includes curated descriptions of each RNA family. These descriptions are quite general as they describe the function across all organisms in which the family is found. RNAcentral imports data from other resources and as of release 22 contains data from 52 other resources of which 12 provide curated data. RNAcentral has previously made efforts to connect users with the relevant literature with the development of the LitScan tool (described in detail in the Sentence Acquisition section below) to explore the EuropePMC API and extract citations and relevant sentences from the literature. However LitScan still lacks a way to provide a coherent and comprehensive overview of an RNA. As a comprehensive source of information on ncRNA, the RNAcentral database is a natural location for the development of tools to more easily connect users with the ncRNA literature.

In this work, we apply a tool based on GPT4, developed by OpenAI, to produce automated summaries for a large number of ncRNA genes. Summaries are generated from sentences mentioning ncRNAs extracted from the literature, and displayed on the RNAcentral website. We detail our approach to sentence acquisition by exploring the EuropePMC API to allow the extraction of relevant passages. These snippets are then passed through a pipeline of selection, summarisation, automated checking, and automated refinement when necessary, which we named LitSumm. The output of this is 4,618 summaries detailing the literature relating to approximately 28,700 transcripts. A randomly selected subset of 50 summaries representative of RNA type and context size are manually evaluated by four expert raters.

# Methods and Materials

## RNA selection

To keep costs and computation size within reasonable limits we focus on a subset of RNAs of broad interest to the community. We include RNAs contributed by HGNC (10), miRBase (11), mirGeneDB (12) and snoDB (13). Within these, we identify primary identifiers and aliases as supplied by the source database.

A large fraction of the RNAs we consider are microRNAs (miRNAs) that are associated with a large corpus of scientific literature. Many of these are referred to by identifiers that are not organism specific such as ‘mir-21’. Having non-specific identifiers leads to a very large number of papers that must be summarised, across a diverse range of organisms; this can lead to confusing or inaccurate statements about the function on a miRNA in a given organism when the function was actually observed elsewhere. More recently, identifiers including an indication of the species have become more common, in this case for example ‘hsa-mir-21’ for the human specific miRNA. The difference in the number of papers discussing these identifiers is enormous. To ensure the specificity of summaries, we restrict the IDs used to generate summaries of miRNAs to only those specific to a species. The exception to this rule is for human miRNAs coming from HGNC, which often have the identifiers like ‘MIR944’, and are included in the set of ncRNAs we summarise.

## Large Language Models

Large Language Models are a class of machine learning models that have very large numbers of parameters, hundreds of billions is common, and are adept at predicting the most probable next token given an input sequence. LLMs are built on the transformer architecture (14). This architecture imposes several limits, first it is expensive in memory and compute to operate on large amounts of text. This imposes a limit on the amount of text, called a context length. Secondly, LLMs do not operate on words, but instead tokens. This means, context lengths are always given as the number of tokens that can be fed into a model; a helpful rule of thumb is that a token is approximately 0.75 words, so a 4,096 token context would be approximately 3,000 words (15).

In this work, GPT4-turbo (https://platform.openai.com/docs/models/gpt-4-and-gpt-4-turbo), an LLM provided by OpenAI is used. Specifically, we use the gpt-4-1106-preview model, through the OpenAI API. The primary parameter controlling the text generation is temperature, $T$, which alters the sampling distribution of the next token; $T=0$ would make the model only choose the most likely next token, while higher values allow the model to explore the distribution of next tokens. We use a relatively low $T=0.1$ (default $T=1$), a balance between determinism and flexibility to rewrite parts of the context into a coherent summary. Low $T$ also reduces the likelihood of model ‘hallucinations’, a common problem where the LLM will invent facts (16).

Two other parameters used to control the generation of the model are the presence and frequency penalties. These alter the sampling distribution by adding a penalty to tokens already present in the text, to reduce repetition. They can also be used to encourage reuse by giving negative values. We use a presence penalty of -2 in the initial summary generation call, to ensure the model re-states tokens from the context in the summary, but with a frequency penalty of 1 to avoid repetition. All operations involving the LLM are abstracted using the langchain python library (https://github.com/hwchase17/langchain).

## Sentence Acquisition

To gather what is being said about an RNA in the literature, we explore the EuropePMC API using a query designed to find articles discussing non-coding RNA while minimising false positives. The query used is ‘query=("<RNA ID>" AND ("rna" OR "mrna" OR "ncrna" OR "lncrna" OR "rrna" OR "sncrna") AND IN_EPMC:Y AND OPEN_ACCESS:Y AND NOT SRC:PPR)’, with the collection of terms in parentheses aiming to filter out false positives that mention the ID but not a type of ncRNA; the query also explicitly requires open access and excludes preprints. We restrict this search to the open access subset at EuropePMC such that we can access and re-use the full text, aside from this no other restrictions are placed on the articles we retrieve or use. RNAcentral’s comprehensive and regularly updated collection of cross-references between RNA resources enables us to identify papers that refer to the same RNA using different names or identifiers.

Once articles about an RNA have been identified, the full text is retrieved and searched to 1) validate that the ID is mentioned in the article and 2) extract sentences that mention the ID. The identified article PMCIDs and contained sentences are stored in a database at RNAcentral. The results of this can be seen on RNAcentral, where the tool is referred to as LitScan (https://rnacentral.org/help/litscan) and has an interface allowing users to explore the results (for example: https://rnacentral.org/rna/URS000075D66B/9606?tab=pub). In this work, we use LitScan as a source of statements about RNAs which can be used to provide an overview of the literature about them.

## Sentence Selection

Not all ncRNAs are studied equally. For many, we know about their existence only because they have been sequenced and deposited in sequence archives such as ENA (17); for these RNAs, we have no papers to summarise. A significant subset of RNAs appear in only a few articles where their existence is established, and occasionally some aspect of function, localisation or other information is determined. To ensure a reasonable amount of information for the LLM to summarise, we restrict the lower bound of sentence count to 5. These 5 sentences could come from a single paper, which allows summarisation of single papers that present the only source of information about an RNA.

Above this threshold, there are two factors driving the selection of sentences from which to summarise: context length and information coverage. For LitSumm, we restrict ourselves to a 4,096 token context, and impose a limit of 2,560 tokens (approximately 1920 words) in the context to allow for prompting and revisions. We limited ourselves to a 4,096 token context for two reasons: feasibility of downstream finetuning of an open LLM; and cost - using the full 128k token window of GPT4 would be prohibitively expensive, since API calls are charged per token. The limit of 2,560 is arrived at by considering the length of output summaries (~255 tokens), prompts (60-140 tokens), and the number of tokens required to send a summary for revision (~1,150) for a random subset of RNAs. For the majority of ncRNAs, the total available sentences fall within this context limit, so no selection is applied beyond the 5 sentence lower limit.

For some ncRNAs such as well studied miRNAs (e.g. hsa-mir-191) and snoRNAs (e.g. SNORD35A) among others, totalling 1704 RNAs, we find too many sentences to use them all, meaning a selection step is necessary. To select sentences, we apply a topic modelling approach (18). We used the SentenceTransformers package (19), with the pretrained 'all-MiniLM-L6-v2' model which embeds each sentence into a 384 dimensional vector, the dimensionality of its output layer. Then, the UMAP dimensionality reduction technique (20) is applied to reduce the vector dimension to 20, and the HDBSCAN clustering algorithm (21) produces clusters of similar sentences. While 384 is not a particularly high dimensionality, it has been shown that reducing dimensionality significantly improves clustering performance across a variety of tasks (22); we choose a dimensionality of 20 such that we can use the fast_hdbscan python library (https://github.com/TutteInstitute/fast_hdbscan), for which 20D is the maximum recommended dimensionality. Cluster exemplars were sampled in a round-robin fashion until the context was filled to ensure a broad coverage of topics. In the case where all exemplars did not fill the context, sentences were sampled from the clusters themselves in the same round-robin way.

An important minority of ncRNAs are very heavily studied. These include ncRNAs like XIST, MALAT1 and NEAT1, each of which appear in thousands of articles. In these cases, our selection technique still results in too many tokens, so we apply a greedy selection algorithm to the cluster exemplars. An exemplar is selected in the largest cluster, then the vector embedding is used to calculate the similarity to all exemplars in other clusters. The exemplar least similar to the selected exemplar is selected, and the process continues by evaluating the distance from all selected exemplars. The process repeats until the context is filled.

For some RNAs, there were too many sentences to use all, but not enough to apply the topic modelling approach. In this case, the sentences were sorted in descending order of tokenised length, and the first k-sentences were taken such that the context was filled.

In all cases, no criteria are applied to the selection of sentences to use in building the summary beyond them having a direct mention of the RNA being summarised.

## Prompts

One of the most critical criteria for a scientific summary is that it contains only factual information. Additionally, tracing the provenance of statements in the summary is important for verifiability. We have designed a chain of prompts through iterative refinement on a subset of examples with these objectives in mind. The first prompt generates the summary, and if there are problems, subsequent prompts attempt to guide the LLM into rectifying them.

The first prompt is shown in Figure 1, panel A.

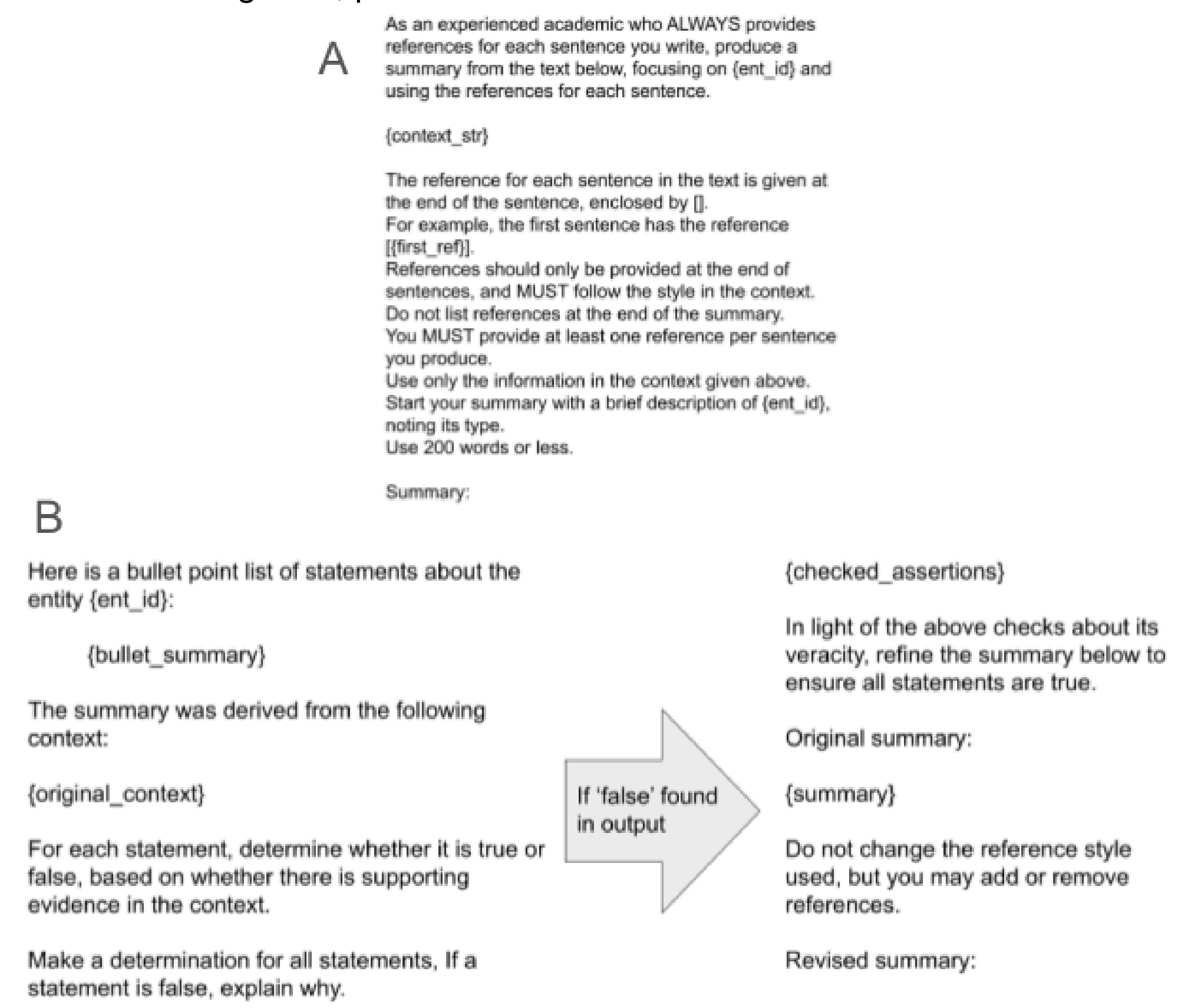

*Figure 1: A: The initial prompt used to generate a first pass summary from the generated context. Variables are enclosed in {} and are replaced with their values before sending the prompt to the LLM. B: Prompts used for the self-consistency checking stage including inaccurate statement detection and revision. All prompts are reproduced as plain text in Supplementary Materials A.*

Here, the model is instructed several times to use references, and the style of reference desired, with an example. The LLM is further instructed not to use 'external sources'; this aims to stop the LLM inserting any facts that are not present in the context. While these facts could be accurate, there is no way of finding out where they come from, and they risk being inaccurate, which we try to avoid at all costs. These instructions, combined with the sampling parameters reduce the likelihood of the model inserting facts not present in the context.

After the summary is returned from the model, references are evaluated. This consists of five checks, any one of which can trigger a re-generation of the summary. The five checks are:

1. Adequacy of references - are there enough references for the number of sentences in the summary? We require at least 0.5 references per sentence.
2. Formatting of references - We require the model to cite sentences by using PubMed Central identifiers [PMCXXXXXX].
3. Realness of references - Are all the references in the summary present in the context? This should catch cases where the model has invented a PMCID.
4. Location of references - references should be at the end of sentences usually. This is intended to stop the model from putting all references at the end of the summary and not indicating which statement comes from which reference.
5. Number of references per instance - this check catches the model putting many PMCIDs into a single pair of brackets, which is undesirable for the purposes of provenance checking. We require no more than 50% of the total number of references in any given citation.

Each check has a specific 'rescue' prompt which is applied when the summary makes the particular mistake. There are four of these, shown in Figure A1 in the supplementary material. To keep costs and computation time within four hundred dollars and a total run-time of 1 day, a maximum of 4 attempts are given to produce a summary. If the summary is still not produced after these attempts, it is flagged as potentially problematic. Provided the summary passes the check within the limit of 4 attempts, it continues to the next stage.

Once all reference based checks have passed, the accuracy of the summary is evaluated. To do this, the summary is broken into a bulleted list, and provided alongside the original context. The model is instructed to state whether each bullet is true or false based on the context, and to find the support in the context. Importantly we not only ask for a true/false, but also ask for an explanation of why. When a summary contains a misleading or false statement, the output of this step, along with the summary, is fed back to the model which is instructed to amend the summary accordingly. These two steps combine approaches to LLM self-fact checking (23) and chain-of-thought prompting (24). In combination, these improve summaries. The prompts used in these stages are shown in Figure 1, panel B.

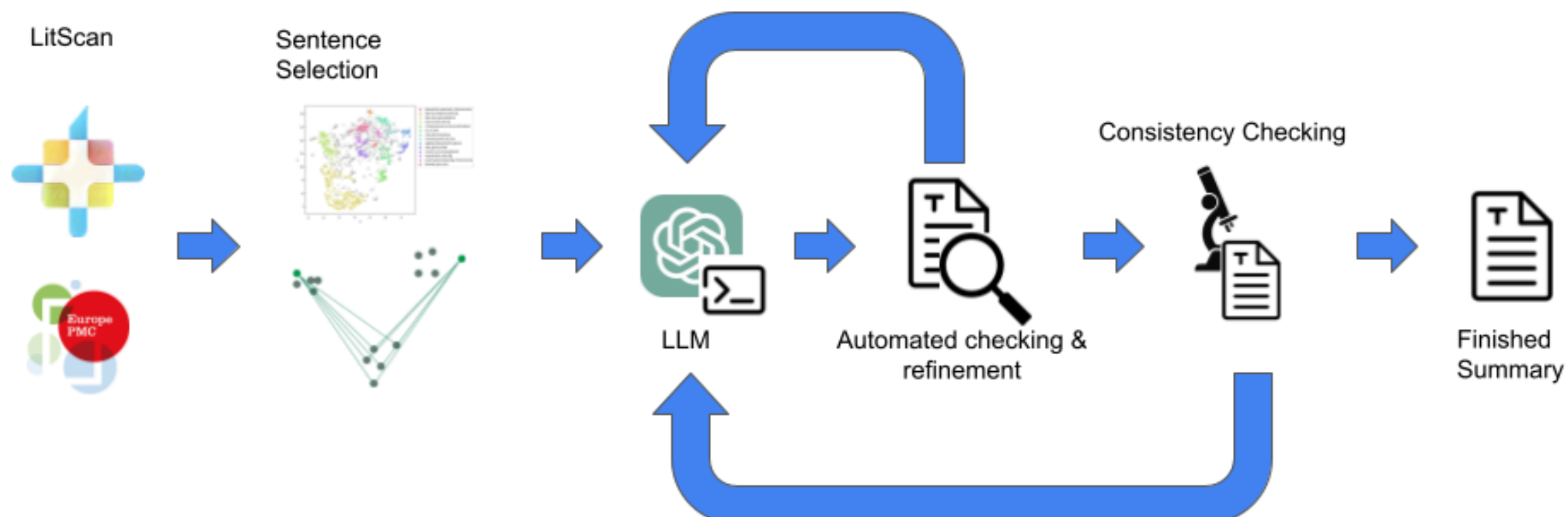

*Figure 2: A flow diagram of the whole LitSumm tool. Information from the EuropePMC API flows from the left to the right, through a sentence selection step before several rounds of self checking and refinement. Finished summaries are written to disk before being uploaded to the RNAcentral database en-masse.*

Once this stage is complete, the summary is given a final reference check, and if successful the summary is considered finished. An overview figure of the whole LitSumm tool is shown in Figure 2.

## Human and Automated Assessment

To evaluate the quality of the output, a subset of the summaries was assessed in parallel by four reviewers. These reviewers were chosen from the coauthors and were specifically selected to represent diverse academic backgrounds, including expertise in data curation, RNA biology, and machine learning. A subset of 50 summaries was randomly selected, stratified by context length, and loaded into a web platform to provide feedback. Figure B1 in supplementary materials shows a screenshot of the web platform used. The distribution of summaries over RNA types is shown in Table 1.

*Table 1: Distribution of RNA types for the 50 summaries reviewed. Since we primarily sampled according to the total number of tokens available for each RNA, without considering the type, this distribution broadly follows the distribution of RNA types in the whole dataset.*

| RNA type | Count |
|---|---|
| lncRNA | 23 |
| pre_miRNA | 11 |
| miRNA | 10 |
| snoRNA | 4 |
| other | 2 |

Summaries were presented alongside the context from which they were generated to allow the raters to evaluate the claims made in the summary. The ratings were given on a 1-5 scale based on the rubric shown in Table B1 in supplementary materials. Briefly: a rating of 1 would indicate multiple serious problems with a summary (e.g. fake references, inaccurate statements, etc); 2 indicates at most two misleading/incorrect statements or one serious error; 3 indicates an acceptable summary with at most one minor misleading/incorrect statement; 4 indicates a summary with no incorrect/misleading statements but with other problems such as poor flow; and 5 would indicate an excellent summary (all statements referenced and true, good flow, etc). All summaries rated 3 and above must have correct, adequate references. Raters were asked to score a summary based only on the information in the context, not using any extra information from the linked articles, or their own knowledge. We also provided a series of tickboxes designed to identify particular failure modes, these are also shown in Table B2 of the supplementary materials.

Raters 0-2 were involved in the design of the rubric, and had the same training with the tool. Experience in RNA science differed between the raters, with Raters 0 and 1 being more experienced than Rater 2, and Rater 0 being a professional curator. Rater 3 was not involved in the development of the summarisation or assessment tools, and received written training to use the rubric before completing their rating session.

# Results

We focused on RNAs from authoritative databases including HGNC, miRBase, mirGeneDB, and snoDB, with particular attention to unambiguous identifiers to ensure accurate functional attribution. From these databases, an initial set of 4,618 RNA identifiers were selected for summarisation. Our pipeline extracted relevant sentences from open access papers in EuropePMC using RNA-specific search queries. Representative sentences were selected using topic modelling and sampling to maintain information diversity. These selected sentences were then processed through GPT-4-turbo using a prompt chain that enforces factual accuracy and proper citation. Each generated summary underwent multiple rounds of self-consistency checking and refinement, validating both reference formatting and factual accuracy against the source material. A subset of the final summaries were evaluated by a panel of experts across multiple domains including RNA biology, data curation, and machine learning.

The 4,618 RNA identifiers selected for summarisation represent a coverage of approximately 28,700 transcripts and 4,605 unique RNA sequences in RNAcentral, and approximately 177,500 papers containing the identifiers. The distribution of RNA types is shown in Figure 3.

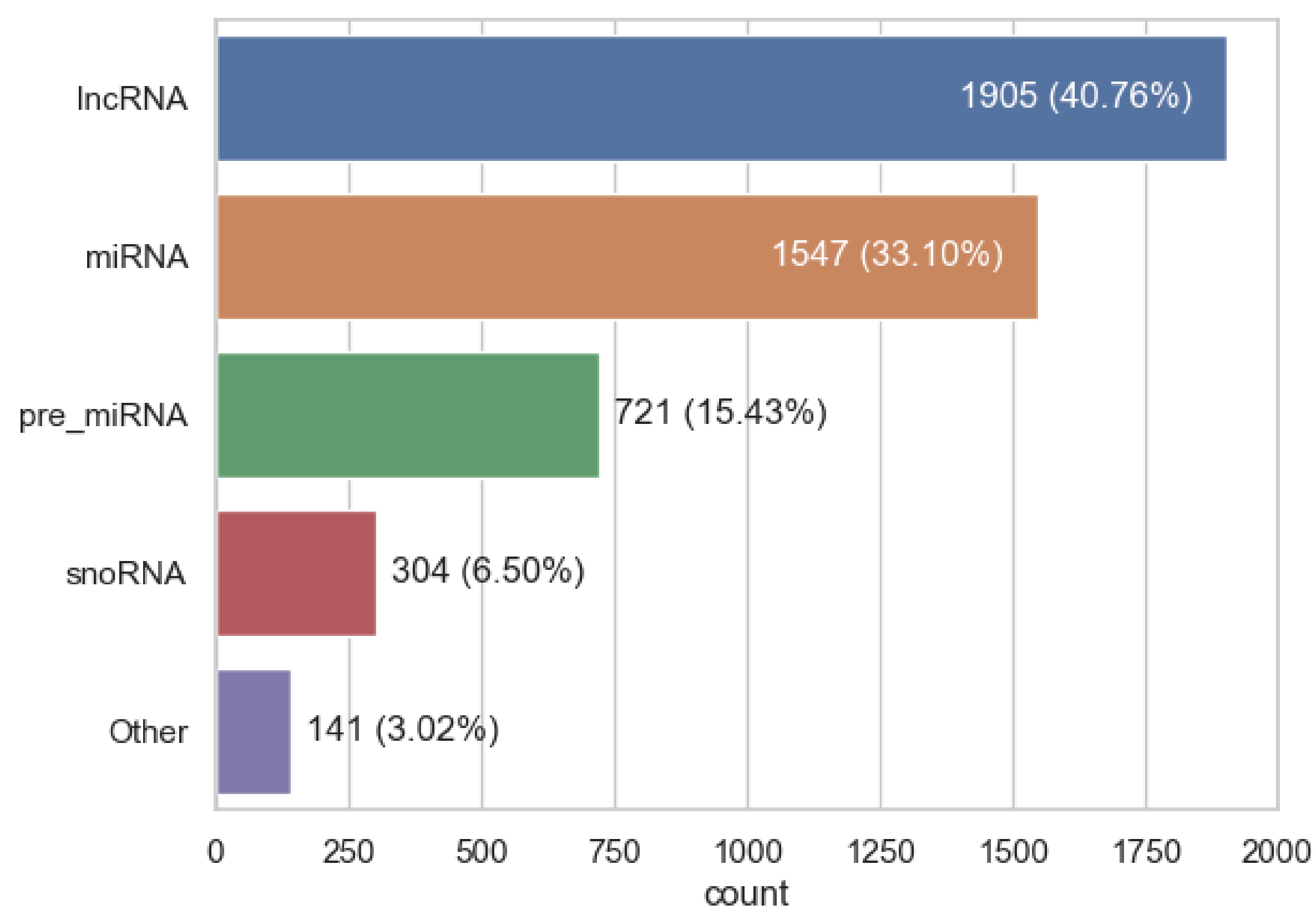

*Figure 3: The distribution of RNA types selected for summarization.*

The majority of RNAs come from the RNA type specific databases miRBase, mirGeneDB and snoDB, who provide miRNAs and snoRNAs; all lncRNAs come from HGNC and are therefore only those found in humans. The small number of ‘other’ type RNAs are from HGNC including for example rRNAs, RNAses and some RNAs with imprecise type labels such as the generic ncRNA. As expected from the chosen databases, the majority of the RNAs selected are human, with non-human RNAs coming primarily from miRBase and mirGeneDB.

The full generation process for each summary, including the automated checking, consistency checking and all revisions took on average 29 seconds and cost $0.05. These values are estimated from the total time to generate all 4618 summaries, and the total cost across the generation period as recorded by OpenAI for billing. An example summary is shown in Figure 4, and all summaries can be browsed by going to the RNAcentral website and searching ‘has_litsumm:”True”’ (https://rnacentral.org/search?q=has_litsumm:%22True%22). We also make the entire dataset of contexts and their summaries available online at https://huggingface.co/datasets/RNAcentral/litsumm-v1.5.

LINC02499, a long intergenic non-coding RNA (lncRNA), has been observed to be significantly downregulated in hepatocellular carcinoma (HCC) tissues compared to adjacent non-tumor tissues, and its decreased expression is associated with poorer overall survival in HCC patients [PMC9775998], [PMC8831247], [PMC9720162]. This lncRNA is one of the top-ranking lncRNAs affecting the expression of cytochrome P450 family members CYP3A5 and CYP3A7, with a notable positive association with CYP3A expression [PMC9775998]. Despite its association with HCC prognosis and cytochrome P450 enzyme expression, the mechanisms by which LINC02499 regulates these processes remain to be elucidated [PMC9775998]. LINC02499 has been identified as a potential prognostic biomarker for HCC due to its significant correlation with clinical factors and overall survival, as well as its inclusion in prognostic models for HCC patients developed using LASSO Cox analysis [PMC9720162], [PMC8167205]. Furthermore, it has been shown to inhibit the proliferation, migration, and invasion of hepatocellular carcinoma cells in vitro [PMC9720162], suggesting a potential role as a tumor suppressor. However, further research is necessary to fully understand its function and clinical utility in HCC diagnosis and prognosis [PMC7812168].

*Figure 4: Example summary generated by the tool. This example is a lncRNA, examples for other RNA types can be found in supplementary materials, appendix C.*

Automated checking, primarily of reference adequacy and accuracy identified problems in 2.1% of summaries which were adequately rectified within the four allowed revisions in 76% of cases, meaning overall 99.5% of summaries passed our automated checks. The self-consistency check identified problems in 17.3% of summaries, which were rectified in 51% of cases giving an overall pass rate of 91.5%. The pass rates at each stage are shown in Table 2.

*Table 2: Pass rates of the automated checking and self-consistency checking stages in the LitSumm pipeline. The table shows the percentage and number of summaries that pass each stage, including revision stages.*

| **Failure mode** | **Pass rate** | **Number of passing summaries** |
|---|---|---|
| References - first pass | 97.9% | 4,519 |
| References - after revision | 99.5% | 4,594 |
| Self-consistency - no problems found | 82.7% | 3,820 |
| Self-consistency - no problems after revision | 91.5% | 4,226 |

An example of the type of error identified and rectified by the consistency check is shown in Figure 5.

- CTBP1-DT is a long non-coding RNA (lncRNA) whose expression is regulated by upstream open reading frames (uORFs) and internal ribosome entry site (IRES) regions within its 5′ untranslated region (5′UTR) [PMC9371908]

True. The context provided indicates that the uORFs and IRES regions within the 5′UTR of the CTBP1-DT gene are involved in the regulation of translation, which is consistent with the statement.

- These elements within the 5′UTR are crucial for the DNA damage-induced upregulation of CTBP1-DT translation [PMC9371908]

False. The context mentions "DNA damage-induced DDUP translation," not CTBP1-DT translation. There is no direct evidence in the provided context that these elements are crucial for the upregulation of CTBP1-DT translation specifically in response to DNA damage.

- Additionally, the expression levels of CTBP1-DT are influenced by the transcription factor ETV5, with changes in ETV5 levels either through knockdown or overexpression affecting CTBP1-DT expression [PMC8660778]

True. The context supports this statement by mentioning that the expression of lncRNA CTBP1-DT was measured following ETV5 knockdown or overexpression, implying that ETV5 levels have an effect on CTBP1-DT expression.

- This suggests a regulatory mechanism where ETV5 may play a role in modulating CTBP1-DT levels in response to DNA damage [PMC8660778].

False. The context provided does not mention DNA damage in relation to ETV5's role in modulating CTBP1-DT levels. The statement extrapolates beyond the provided evidence, which only indicates that ETV5 affects CTBP1-DT expression, not specifically in response to DNA damage.

*Figure 5: Example output of the veracity checker. In this case, CTBP1-DT presents two sentences validated as TRUE and two FALSE sentences. The offending sentences have been removed by the model in the final summary.*

Human evaluation was carried out for a subset of randomly sampled RNAs. These RNAs cover the full range of context sizes and ncRNA types. The subset consists of 50 RNAs in total, for which three raters assessed quality. 21 of these 50, covering only miRNA, were also scored by an expert in miRNA (Rater 3). The human rating, on a scale of 1-5 with 5 being excellent and 1 indicating the presence of some serious failure is shown for all raters in Figure 6.

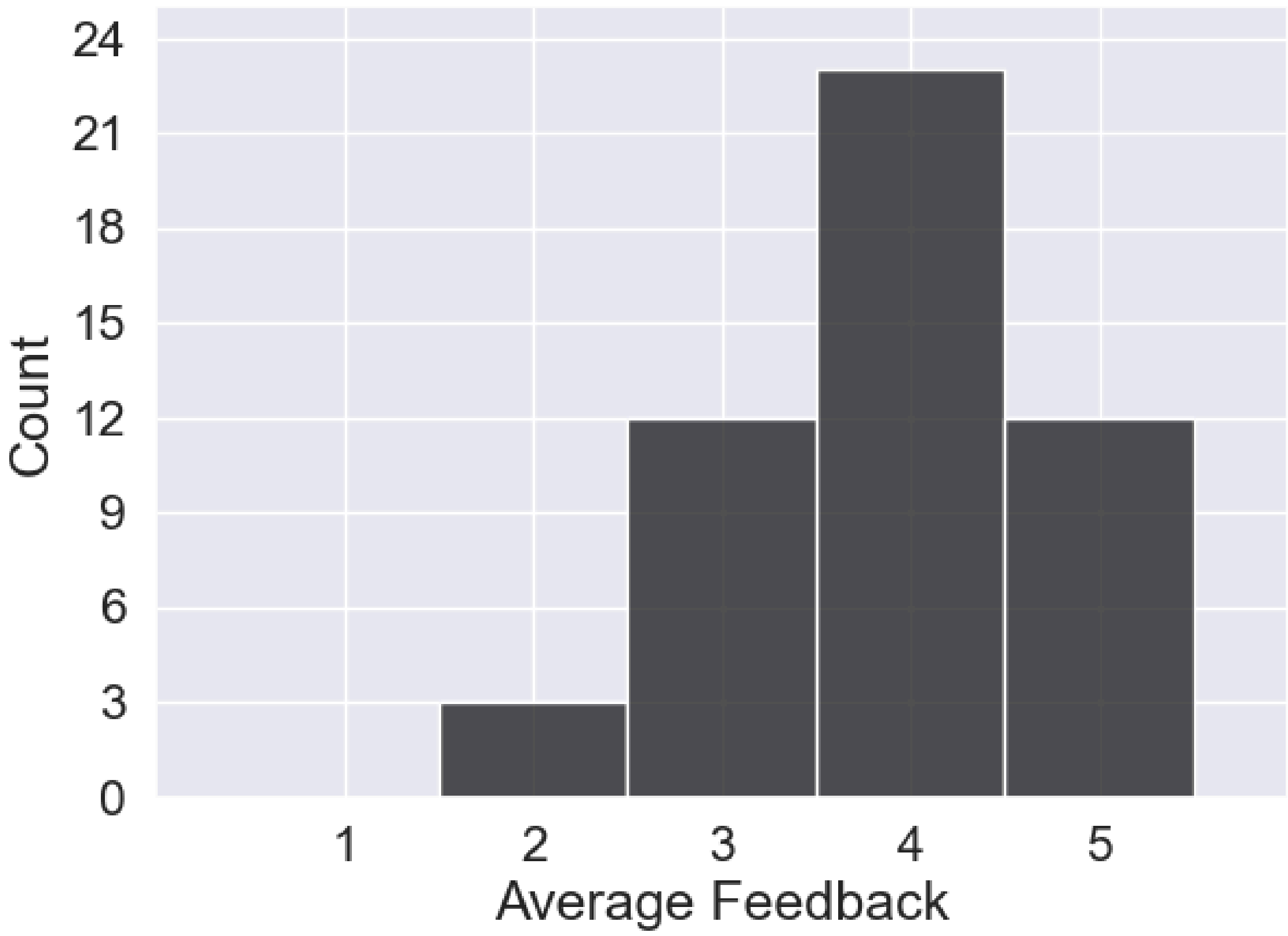

*Figure 6: The average rating per summary across all raters. Note that Rater 3 gave scores only for a subset of 21 miRNAs.*

From the human ratings, 94% of summaries were rated good or excellent. In the majority of cases where a summary was rated inadequate (score of 2 or less) the problem identified by the raters had to do with poor synthesis of facts from multiple sources not caught by the automated consistency check, or reference misattribution, where a reference for a given sentence does not match the information content, or is irrelevant. Reference errors are penalised strongly in the marking rubric, as are misleading statements. LLMs are known to struggle to accurately combine facts across different documents (25). This has been observed in previous studies of multi-document summarizations and may be connected to input construction. A summary of the failures identified is shown in Table 3, and the best and worst rated summaries per RNA type are shown in Supplementary Materials Appendix C.

Summaries rated inadequate by human inspection are retained, but not displayed on RNAcentral. Similarly, where problems were identified by automated checks, e.g. failing to insert correct references, or automatically detected self consistency errors, the summaries are retained but not displayed. This means that from a total of 4,605 unique RNA sequences, 4,602 have a displayed summary on the RNAcentral website

*Table 3: Reasons given for poor rating in those cases where a rating less than 3 was given.*

| Rater | Total (out of 50 summaries) | Reference Formatting | Hallucination/false info | Other |
|---|---|---|---|---|
| Rater 0 | 9 | 4 | 9 | 1 |
| Rater 1 | 5 | 0 | 3 | 2 |
| Rater 2 | 7 | 3 | 3 | 3 |
| Rater 3 | 3/21 | 0 | 1/21 | 4/21 |

~~Our analysis of the literature summaries revealed several notable issues with the performance of the LLM. Firstly, the model struggled with synthesising information from diverse sources, leading to unwarranted inferences and a lack of support for its claims.~~

Additionally, we observed instances of reference misattribution, where unrelated or incorrect references were added to sentences, undermining the ease of verification of the summaries. Furthermore, the model exhibited a tendency to over-infer and make unwarranted connections, often overstating the significance of findings, such as prematurely identifying biomarkers. We also noted cases of unsupported expansions, where the model introduced information that was not supported by the original text; for example expanding DFU as 'Diabetic Foot Ulcer' when the provided context made no mention of diabetes, feet or ulcers. Finally, issues with flow and coherence were evident, including the inclusion of internal instructions and inappropriate recommendations. These findings highlight the need for continued refinement and evaluation of LLMs in generating accurate and reliable summaries of scientific literature.

Summaries are available on RNAcentral by exploring results of the search 'has_litsumm:"True"' (https://rnacentral.org/search?q=has_litsumm:%22True%22), or by exploring the "Literature Integration" facet and selecting "AI generated summaries". We also make the entire dataset of contexts and their summaries available online at https://huggingface.co/datasets/RNAcentral/litsumm-v1.5. Summaries will be updated with each RNAcentral release by re-running the LitSumm pipeline, including re-running LitScan to identify new literature; future releases may move to the use of a local model. Previous versions of the summaries will be versioned and made available, including metadata on the generation, such as the LLM model and sentence selection techniques used. Future work will include moving to a fine-tuned local model, which will allow expansion to more ncRNAs, and a more even taxonomic coverage without significant additional cost.

# Discussion

In this work we present an application of LLMs to perform literature curation for ncRNA. We show that a pipeline with a series of automated checks and carefully designed prompts can produce high-quality literature summaries. We also demonstrate techniques to minimise untrue information, and ensure high quality referencing in the summary.

Human ratings of a representative subset of the summaries generated have been collected, and show that the majority of summaries are of high or very high quality, with a small number of common failure modes. The identified failure modes primarily fall into two categories - relating to referencing, and relating to information synthesis/inference from multiple sources.

The size of the fully annotated set is small, at only 50 summaries, but thorough, with three of the raters rating all 50 summaries. This is in line with the sizes of other multi-rated summarization datasets (26), but with a greater overlap (100% vs 20% in Wang et al.). Collecting multiple ratings for each summary improves our sensitivity to nuanced errors at the cost of coverage. The failure modes we identified were consistent across instances, and picked up by all raters; however, more nuanced errors were identified in context-specific cases and not picked up by all raters. We are confident that our ratings identified the nuanced errors made by the LLM, but given the small coverage, identifying a clear pattern to these errors is difficult. It may be preferable to allocate rating time differently to maximise coverage with some minimal overlap to assess consistency.

In this work, the human ratings have been performed by four of the co-authors of the paper for all summaries. Having authors rate summaries introduces a conflict of interest in that the authors have a vested interest in the ratings being ‘good’. To mitigate the potential for bias, we employed a grading rubric and multiple rating for all summaries, and further asked an external expert to rate summaries in their own field of expertise (also a co-author). We believe the bias is minimal, evidenced by the low ratings for poor quality summaries given by all raters. In future it would be preferable to have additional independent ratings.

Summarisation is a task which has been approached by language models previously, such as the T5 architecture (27). While these models do perform well on summarisation and other tasks, they are not as general purpose as a modern, instruction tuned LLMs which are often equally adept at summarisation. As such, basing the LitSumm architecture on a single driving model simplifies the tool. LLM driven summarisation has been done in several other fields. For example, Joachimiak et al. developed a similar tool, SPINDOCTOR, which is used to generate a summary from gene descriptions; the summary is then used in a gene enrichment analysis (5). Joachimiak *et al.* evaluate the results of their gene enrichment against standard tools and find their method is comparable, though it misses some important terms. SPINDOCTOR differs from LitSumm in that the input is a set of genes known to be enriched in an experiment, and the output is a summary of their commonalities, whereas LitSumm produces a broad overview summary about a single RNA from many literature-derived statements. SPINDOCTOR also does not need to assess the consistency of their summary with the context from which it is generated, and do not give the provenance of statements, since their input is human-derived.

One field in which similar considerations have to be made is medicine, where the accuracy and provenance of statements is paramount. Shaib *et al.* evaluate GPT 3 for the summarization and synthesis of many randomised controlled trial reports. They find that while the LLM produces coherent summaries, it often fails to synthesise information from multiple sources adequately, and may be over-confident in its conclusions (28). In our evaluation we find similar failure modes, where the model misunderstands statements where it tries to synthesise information from more than one source.

A key aspect of our pipeline is the use of self consistency checking and revision using chain-of-thought prompting. These two concepts have been applied in other contexts, such as question answering over documents (29), but have yet to be applied to literature curation. Despite our best efforts to reduce hallucinations and ensure wholly factual summaries, around 17% of cases still have some problems, indicating the need for consistency checking. Feeding the output of the self checking back into the model reduces this to 8.5%, which is encouraging, but also indicates the need for human intervention in this complex field where LLMs still struggle to fully comprehend scientific literature. In particular, the consistency check developed here is not effective at identifying inferences made by the LLM that are incorrect, because there is 'indirect' support in the context. There is also the possibility of 'feedback hallucination' in which the generated instructions to rescue a summary contain a hallucination, which if not checked will be inserted by the LLM as it follows its own instructions. As LLMs become stronger, and have more advanced reasoning capabilities, this will become an increasingly problematic failure mode; the detection of errors of this sort is an area of active research in the NLP field generally.

We investigated the observed errors in the produced summaries, finding that the majority of poorly rated summaries share several common problems. Namely, we observe incorrect reference attribution, the inclusion of irrelevant details while missing important information, and statements unsupported by the context as the most common problems. More details of this error analysis can be found in Supplementary materials Table B2 and Appendix C.

Another limitation relates to the literature itself, and here is primarily seen with miRNAs. Many gene names or identifiers are ambiguous in that they can be used to refer to multiple organisms, or may conflate the mature product and precursor hairpin. We have restricted ourselves to species specific IDs (e.g. hsa-mir-126), meaning the generated summaries should be consistent and limited to a single organism, but a significant fraction of the literature does not use these IDs. Thus, we are missing information. We could use a broader set of identifiers, but then we must be able to distinguish which species is being discussed in each paper. There are ways this could be addressed - for example using the ORGANISMS database (30) to identify which organism a given article is about and then use this information to produce organism specific summaries, despite the usage of non-specific terms. However, the accuracy of such resources is questionable, meaning we do not know which organism a paper discusses at present. We leave this problem as future work.

Other automated assessment methods have been developed that use another LLM to give a rating to the output of an LLM. This can be done with very little guidance as in single answer grading, described by (31), where GPT4 is simply asked to grade output, or in a much more guided and structured way as in Prometheus (32) where a gold standard answer and marking rubric are provided to the model. While LLM rating approaches have not been applied here, such a tool would be valuable to ensure the quality of summaries without extensive human curation. However, the development of a suitable rubric is not straightforward; we plan to approach this problem in a future work.

One limitation that will be difficult to address is the openness of literature. Our sentences come from the open-access subset of EuropePMC. While this data source is growing as more authors publish open-access, it still does not allow access to the majority of knowledge, particularly that from earlier decades. Many knowledgebases make extensive use of closed-access literature in their curation; their primary concern is the quality of the information being curated, not the availability of the information, therefore the open access status of a paper is not an impediment to its being curated. However, the inability of this tool, and those which will doubtless come after it, to use closed-access literature does highlight the need for authors, institutions and funders to push for open access publication with a permissive licence for reuse.

Often there is too much literature available to feed all of it into an LLM to generate a summary. Recently, LLMs have been getting considerably larger context sizes, for example GPT4 can now accept up to 128k tokens. However, this is unlikely to be a solution in itself; LLMs do not attend to their entire context equally (33), and having a larger context and expecting the LLM to use it all is unlikely to work, though some recent work has shown that this may be soluble (34). In this work, topic modelling is used to reduce the amount of text to be summarised. This introduces problems related to the context construction that lead to inaccurate sentences being generated by the LLM. Worse, the automated fact checking is blind to this type of failure, due to there being ‘evidence’ in the context which supports the inaccurate sentence. Therefore, we would recommend against the use of topic modelling alone to generate input context for an LLM, since it likely introduces more problems than alternative approaches such as vector store based retrieval augmented generation (RAG). A better approach may be to decompose the summary into sections and apply a RAG (35) approach to each in turn by applying semantic search for only passages about, for example, expression.

The field of LLM research is moving extremely rapidly and we expect that significant improvements will be possible in our pipeline simply by adopting newer LLM technology. Our current work is based on GPT4, having been originally developed with GPT3.5; this allows us to see the improvement in LLM technology, which we show in Appendix D with a brief A/B preference test between GPT3.5 and 4, and examination of pass rates through our pipeline. Moving to openly available models could enable future work on fine-tuning the LLM for the biological summarisation task.

# Conclusion

In conclusion, we have demonstrated that LLMs are a powerful tool for the summarisation of scientific literature and, with appropriate prompting and self checking, can produce summaries of high quality with adequate references. Using the tool developed here, 4,618 high quality summaries have been provided for RNAcentral, providing natural language summaries for these RNAs where none previously existed. This is the first step to automating the summarisation of literature in ncRNAs, and providing helpful overviews to researchers.

# Availability

All code used to produce these summaries can be found here: https://github.com/RNAcentral/litscan-summarization and the dataset of contexts and summaries can be found here: https://huggingface.co/datasets/RNAcentral/litsumm-v1.5. Summaries are also displayed on the RNA report pages in RNAcentral (https://rnacentral.org/) and can be explored by searching with the query 'has_litsumm:"True"' (https://rnacentral.org/search?q=has_litsumm:%22True%22).

*Conflict of Interest:* A.B. is an Editor at DATABASE, but was not involved in the editorial process of this manuscript.

## Funding

This project has received funding from the European Union's Horizon 2020 Marie Skłodowska-Curie Actions under grant agreement No 945405; and from Wellcome Trust [218302/Z/19/Z]. For the purpose of open access, the author has applied a CC BY public copyright licence to any Author Accepted Manuscript version arising from this submission. AB and BS are funded by core EMBL funds.